\documentclass[twocolumn,pra,showpacs,showkeys,amsmath,amssymb]{revtex4}
\usepackage{graphicx,amsmath,epsfig,latexsym,color}
\newcommand{\be}{\begin{equation}}
\newcommand{\ee}{\end{equation}}
\begin{document}

\title{Limitations of Nanotechnology for Atom Interferometry}
\author{Alexander D.\ Cronin, Lu Wang, John D. Perreault}
\affiliation{Department of Physics, University of Arizona, 1118 E
4th St. Tucson, Arizona 85721}
\date{\today}


\begin{abstract}
Do van der Waals interactions determine the smallest
nanostructures that can be used for atom optics? This question is
studied with regard to the problem of designing an atom
interferometer with optimum sensitivity to de Broglie wave phase
shifts. The optimum sensitivity to acceleration and rotation rates
is also considered. For these applications we predict that
nanostructures with a period smaller than 40 nm will cause atom
interferometers to perform poorly because van der Waals
interactions adversely affect how nanostructure gratings work as
beam-splitters.
\end{abstract}
\date{\today}
\maketitle

Atom interferometers that are built with nanostructure gratings
have proven their ability to detect small perturbations to atomic
de Broglie waves. Examples of quantities measured with this
technique include: the polarizability of Na atoms \cite{ESC95},
the index of refraction for Na atom waves due to a dilute gas
\cite{SCE95,RCK02}, the strength of atom-surface van der Waals
interaction potentials \cite{PEC05}, and the rotation rate of a
platform \cite{HCL97,LHS97}. Because all of these measurements are
related to interference fringe phase shifts, an important design
goal for atom interferometers is to optimize the sensitivity to
the phase of an interference pattern. This goal was discussed by
\textcite{SCD93} for matter-wave interferometers in general, and
then discussed for atom interferometers that are based on
mechanical absorption gratings by Pritchard \cite{BER97,LHS97} and
also by Vigu\'{e} \cite{CBV99,DML02}. However, none of these
analyses specifically include the effect of van der Waals (vdW)
interactions between atoms and the nanostrucure gratings.  In this
paper we review how phase sensitivity can be maximized by
selecting the open fraction of each grating; then we show how vdW
interactions modify these calculations. Finally, we show how vdW
interactions determine the minimum period of nanostructure
gratings that can optimize the performance of atom interferometers
for inertial sensing.

van der Waals interactions between atoms and material gratings
change the diffraction efficiencies, $e_n$, which we define as the
modulus of the diffracted wave amplitude in the $n$th order as
compared the wave amplitude incident on the grating, \be e_n =
\left|\frac{\psi_n}{\psi_{inc}}\right|. \ee  The effect of vdW
interactions is generally marked by an increase in $e_n$ for $n>1$
and a decrease in $e_0$, but the efficiencies $e_n$ depend
non-linearly on the strength of vdW interactions as discussed in
\cite{gris99, bruh02, GST00, PCS05, CRP04}. In this paper we will
use newly identified scaling laws for $e_n$ to describe how the
vdW interactions affect the interference pattern and the
statistical sensitivity to phase shifts in a three-grating
Mach-Zehnder atom interferometer. The impact of vdW interactions
is particularly important if the grating period is reduced below
100-nm and the atom beam velocity is reduced below 1000 m/s. This
analysis shows how vdW forces set fundamental limits on the
smallest nanostructure features that can be used for atom
interferometry. For example, grating windows that are 20 nm wide
will optimize the sensitivity to rotations or accelerations for an
atom interferometer with 1000 m/s Na atoms and 150 nm thick
silicon nitride gratings.

\begin{figure}[t]
\begin{center}
\includegraphics[width=7cm] {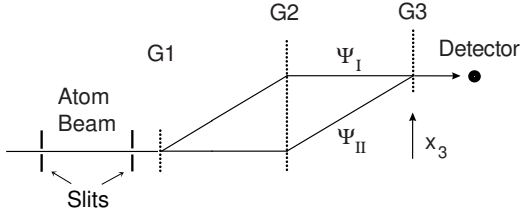}
\caption{Interferometer layout with gratings G1, G2 and G3. The
transverse position of G3 relative to the other gratings is
labeled $x_3$. Two beams with amplitudes $\psi_{\texttt{I}}$ and
$\psi_{\texttt{II}}$ are incident on G3.} \label{fig:layout}
\end{center}
\end{figure}

\begin{figure}[t]
\begin{center}
\includegraphics[width=8cm] {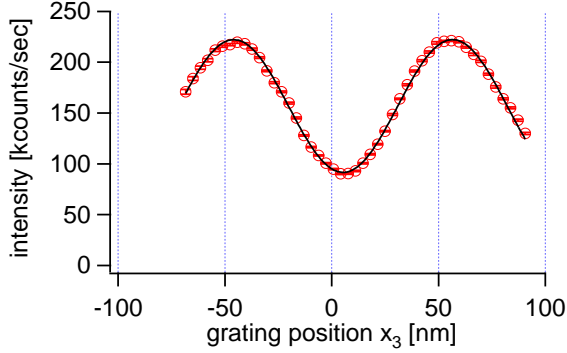}
\caption{Interference fringe data and best fit based on equation 2
with $\langle I \rangle$ = 157,000 counts per second and $C = 0.42$.
A total of 5 seconds of data are shown and the uncertainty in phase
calculated by equation \ref{eq:sigma_phi} is $\sigma_{\phi}= 2.7
\times 10^{-3}$ radians.} \label{fig:fringe}
\end{center}
\end{figure}

The layout of a Mach Zehnder atom beam interferometer is shown in
figure \ref{fig:layout}.  The first grating (G1) serves as a beam
splitter. The second grating (G2) redirects the beams so they
overlap in space and make a probability density interference
pattern $I'(x')$ just before the third grating (G3). Because the
detector in this example is located close enough to G3 so that
diffraction from G3 is not resolved, the third grating simply
works as a mask. The transmitted flux $I(x_3)$ depends on position
of the third grating, $x_3$, relative to the other gratings as \be
I(x_3) = \langle I \rangle \left[1 + C \cos(k_g x_3 + \phi)
\right] \label{eq:Ix3} \ee where $\langle I \rangle$ is the
average transmitted atom beam intensity, $C$ is the contrast
defined as \be C = \frac{ I_{\textrm{Max}} -
I_{\textrm{Min}}}{I_{\textrm{Max}} + I_{\textrm{Min}}}, \ee
$k_g=2\pi/d$ is the wavenumber of the grating, and $\phi$ is the
phase that the interferometer is designed to measure.   An example
of interference fringe data and a best-fit function for $I(x_3)$
based on equation \ref{eq:Ix3} is shown in figure
\ref{fig:fringe}. The values of $\langle I \rangle$, $C$, and
$\phi$ are free parameters in the fit, but the period $d= 100$ nm
in figure \ref{fig:fringe} is determined by the gratings, and the
distance $x_3$ is independently measured with a laser
interferometer as described in \cite{KET91}.

As discussed in \cite{SCD93,LHS97,BER97}, the measured fringe phase,
$\phi$, is predicted to have a statistical variance,
$\sigma_{\phi}^2$, due to shot noise (counting statistics) given by
\be (\sigma_{\phi})^2 \equiv \left\langle (\phi - \langle \phi
\rangle )^2 \right\rangle = \frac{1}{C^2 N}\label{eq:sigma_phi} \ee
where $N$ is the total number of atoms counted. To minimize the
uncertainty in phase we therefore seek to maximize the quantity
$C^2\langle I \rangle$ which is proportional to $C^2 N$. This can be
achieved by increasing the observation time, increasing the
intensity of the atom beam incident on the interferometer
($I_{inc}$), increasing the quantum efficiency of the detector, and,
depending on the beam collimation,
by increasing the detector size.
Maximizing $C^2 \langle I \rangle/I_{inc}$ can also be achieved by
choosing specific open fractions for the three gratings because the
open fractions affect $\psi_{\texttt{I}}$ and $\psi_{\texttt{II}}$
relative to $\psi_{inc}$. The open fractions are defined as $w_i/d$
where $w_i$ is the window size for the $i$th grating (G1 G2 or G3),
and $d$ is the grating period.  We will show that other factors in
addition to the open fractions such as the strength of the vdW
interaction, the atom beam velocity, and the absolute size of the
nanostructure gratings are needed to completely determine
$\psi_{\texttt{I}}$ and $\psi_{\texttt{II}}$ relative to
$\psi_{inc}$.

The interference pattern just \emph{before} G3 is
\begin{eqnarray}
I'(x') &= &\left| \psi_{\texttt{I}} + \psi_{\texttt{II}} \right|^2 \label{eq:I'(x')amplitudes} \\
&=& \left| e_1^{G1} e_{-1}^{G2} +e_0^{G1} e_1^{G2} \textrm{
}e^{i(k_g x' + \phi)}\right|^2 I_{inc}
\label{eq:I'(x')efficiencies} \\ &=& \langle I' \rangle [1 + C'
\cos(k_g x' + \phi) ] \label{eq:I'(x')cosine}
\end{eqnarray}
where $\psi_{\texttt{I}}$ and $\psi_{\texttt{II}}$ are the wave
functions of the two beams incident on G3, and the notations
$e_n^{G1}$ and $e_n^{G2}$ denote the diffraction efficiency for G1
or G2 into the $n$th order. Equation \ref{eq:I'(x')cosine} looks
similar to equation \ref{eq:Ix3} but the primes indicate the
position, intensity, and contrast of the interference pattern just
before G3.

The intensity transmitted through the third grating, $I(x_3)$, is
related to the intensity $I'(x')$ by \be I(x_3) = \frac{1}{d}
\int_{-w_3/2}^{w_3/2} I'(x_3-x') dx'. \label{eq:int3}\ee Equation
\ref{eq:int3} describes the result of Moir\'{e} filtering of the
interference pattern $I'(x')$ by G3, with the grating acting as a
binary valued (Ronchi-rule) mask. The result of combining
equations \ref{eq:I'(x')cosine} and \ref{eq:int3} to get \be
I(x_3) = \langle I' \rangle \frac{w_3}{d} \left[ 1 + C'
\frac{\sin(k_g w_3/2)}{(k_g w_3/2)} \cos(k_g x_3 + \phi)\right]
\ee lets us identify $\langle I \rangle$ and $C$ in terms of
$\langle I' \rangle$ and $C'$ respectively as
\begin{eqnarray} \langle I \rangle &=& \frac{w_3}{d} \langle I' \rangle,
\label{eq:IIp} \\ C &=& \frac{\sin(k_g w_3/2)}{(k_g w_3/2)}C'.
\label{eq:CCp} \end{eqnarray} Furthermore, the values $\langle I'
\rangle$ and $C'$ are related to the diffraction efficiencies by
equations \ref{eq:I'(x')efficiencies} and \ref{eq:I'(x')cosine},
so that after the third grating $\langle I \rangle$ and $C$ can be
expressed as \begin{eqnarray} \langle I \rangle &=& \left[
\left(e_0^{G1} e_1^{G2}\right)^2 + \left(e_1^{G1}
e_{-1}^{G2}\right)^2 \right] \left[\frac{w_3}{d}\right]I_{inc},\\
 C&=& \frac{2 \textrm{ } e_1^{G1}e_{-1}^{G2}
e_0^{G1}e_1^{G2}}{(e_1^{G1}e_{-1}^{G2})^2 + (e_0^{G1}e_1^{G2})^2 }
\left[\frac{\sin(k_g w_3 /2)}{(k_g w_3 /2)}\right].
\end{eqnarray}

The quantity we seek to maximize, $C^2 \langle I \rangle$, can now
be written in terms of the diffraction efficiencies for G1 and G2
and the open fraction for G3:

\begin{widetext}
\be C^2 \langle I \rangle  = 4 I_{inc}
\left[\frac{(e_1^{G1}e_0^{G1})^2}{ (e_1^{G1})^2 +
(e_0^{G1})^2}\right] \left[\left(e_1^{G2}\right)^2\right]
\left[\left(\frac{\sin(k_g w_3 /2)}{(k_g w_3 /2)}\right)^2
\left(\frac{w_3}{d}\right)\right]. \label{eq:FOMeew} \ee
\end{widetext}

\noindent The figure of merit, $C^2 \langle I \rangle$, in equation
\ref{eq:FOMeew} has been factored into three bracketed terms that
are each determined by only one grating G1 G2 or G3. Each term in
brackets can then be maximized independently. To obtain equation
\ref{eq:FOMeew}, it was assumed that $e_1^{G2} = e_{-1}^{G2}$, which
is verified to be a good approximation by the symmetry of the
experimentally observed diffraction patterns \cite{CEH95,GST00}.

Note that if diffraction from G3 can be resolved, as in the
interferometer built by Toennies \cite{TOE01,TOE05}, then G3 acts
as a beam combiner (not a mask) and $C^2 \langle I \rangle$
becomes a function of the diffraction efficiencies of all three
gratings.   This applies also for phase gratings and has been
considered by Vigu\'{e} \cite{CBV99}, however we will restrict
this paper on vdW interactions to the case of nanostructure
gratings with G3 acting as a mask.   Near field diffraction from
the collimating slits and $C_3$ dependent near field effects from
G3 are explicitly ignored here.

To state the figure of merit $C^2 \langle I \rangle$ in equation
\ref{eq:FOMeew} in terms of the physical dimensions of each grating
the next step is to evaluate the diffraction efficiencies. \emph{If
vdW interactions with the grating bars are ignored} then the
diffraction efficiency for atom wave amplitude into the $n$th
transmission diffraction order is given by \be e_n =
\left(\frac{w}{d}\right) \frac{\sin(n\pi w /d)}{(n\pi w/d)}
\label{eq:sinc} \ee where $w$ is the grating window size and $d$ is
the grating period. Equation \ref{eq:sinc} is valid in the far-field
(Fraunhofer) approximation and gives the familiar sinc$^2(nw/d)$
envelope function for diffraction intensities $I_n = |\psi_n|^2$.

\begin{figure}[t!]
\begin{center}
\includegraphics[width=8cm] {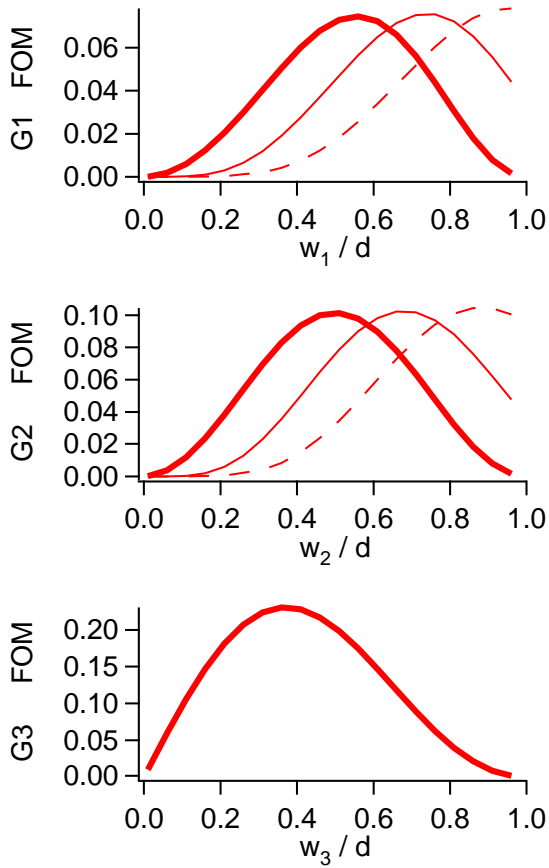}
\caption{The bracketed terms in equation \ref{eq:FOMeew} that
optimize $C^2 \langle I \rangle / I_{inc}$ are plotted as a
function of the open fractions $w_1$, $w_2$, and $w_3$ (Top,
Middle and Bottom). The thick solid curves correspond to $C_3$=0.
The factors are also shown for the cases of $C_3 = 3$ and 30
meVnm$^3$ (thick, thin and dashed lines). The additional
parameters are $v=1000$m/s, $d=100$nm and $\ell=150$nm.}
\label{fig:w123}
\end{center}
\end{figure}

Using equation \ref{eq:sinc} for the diffraction efficiencies we
can write $C^2 \langle I \rangle$ in terms of the open fractions
of the three gratings:
\begin{widetext}
\be C^2 \langle I \rangle  =  4I_{inc}
\left[\frac{(\frac{w_1}{d}\frac{\sin(\pi w_1
/d)}{\pi})^2}{\left(\frac{w_1}{d}\right)^2 + \left( \frac{\sin ( \pi
w_1 / d)}{\pi}\right)^2} \right] \left[\left( \frac{\sin(\pi
w_2/d)}{\pi}\right)^2\right] \left[\left(\frac{\sin(\pi w_3
/d)}{(\pi w_3 /d)}\right)^2 \left(\frac{w_3}{d}\right)\right].
\label{eq:FOMwww} \ee
\end{widetext}
The three bracketed terms in equation \ref{eq:FOMwww} are
functions of $w_1d^{-1}$,  $w_2d^{-1}$, and $w_3d^{-1}$
respectively, and each term can be maximized independently as
shown in Figure \ref{fig:w123}. The open fractions for (G1, G3,
G3) that maximize $C^2\langle I \rangle$ are (0.56, 0.50, 0.37).
The maximum value of $C^2 \langle I \rangle/I_{inc} = 0.0070$ is
obtained when the contrast is $C= 0.67$ and the average intensity
is $\langle I \rangle = (0.015) I_{inc}$. It is noteworthy that a
low value of $\langle I \rangle/I_{inc} \ll 1$ is hard to avoid
because the gratings are, after all, absorption gratings. These
values for $w_1$, $w_2$, $w_3$, $C$, and $\langle I
\rangle/I_{inc}$ that maximize $C^2\langle I \rangle$ reproduce
the results stated in \cite{BER97} and are listed on the first row
of table \ref{tab:C3}.  This concludes our review of how to get
optimum interference fringe patterns (with minimum
$\sigma_{\phi}$) from an atom interferometer built with three
nanostructure gratings assuming vdW interactions are negligible.

Next we will show how vdW interactions modify the best open
fractions for G1 and G2. To predict how the diffraction efficiencies
change as a result of the vdW interaction we will use a numerical
calculation described in \cite{PCS05,CRP04} and summarized here. The
van der Waals potential is \be V(r) = -\frac{C_3}{r^3}, \ee where
$r$ is the distance to an infinite plane and $C_3$ is the \emph{vdW
coefficient}.   The vdW coefficient for sodium atoms and silicon
nitride surfaces has been measured to be $C_3 = 3$ meVnm$^3$
\cite{PCS05} and for helium and silicon nitride $C_3 = 0.1$
meVnm$^3$ \cite{gris99}. The phase shift for atom waves passing
through a slot between two grating bars, as discussed in references
\cite{PEC05,CRP04,PCS05,gris99,GST00,bruh02}, is \be \phi(\xi)=
\frac{C_3 \ell}{\hbar v} \left(\frac{1}{|\xi-w/2|^3} +
\frac{1}{|\xi+w/2|^3}\right) \label{eq:phi}\ee where $\xi$ is the
coordinate inside the grating channel ($\xi = 0$ in the middle of
the window), $\ell$ is the thickness of the grating, $\hbar$ is
Planck's constant divided by 2$\pi$, and $v$ is the atom beam
velocity. Equation \ref{eq:phi} is obtained by assuming
parallel-sided slot walls, neglecting edge effects at the entrance
and exit to the slot, and using the WKB approximation to first order
in $V(r)/E$ (potential over kinetic energy). Despite all these
approximations, equation \ref{eq:phi} has been used (occasionally
with a modification for non-parallel walls) to explain several
experimental observations regarding vdW interactions between atoms
and nanostructure gratings \cite{PCS05,CRP04,PEC05,gris99,bruh02}.

With the  vdW-induced phase shift,  $\phi(\xi)$,  given by
equation \ref{eq:phi} incorporated into the transmission function
for the grating, the diffraction efficiencies in the far-field
approximation are \be e_n = \frac{1}{d} \int^{w/2}_{-w/2} \exp
\left[ ink_g\xi + i\phi(\xi) \right] d\xi. \label{envdW} \ee

By combining equations \ref{eq:phi} and \ref{envdW} and performing
the change of variables $\xi = yd$ we can re-write the
efficiencies in terms of three linearly independent dimensionless
parameters so that $e_n =  e_n(p_1,p_2,p_3)$ is explicitly \be
e_n=
 \int_{-p_2}^{p_2} \exp \left[ip_1 y + \frac{ip_3} {|y-p_2|^3} +
\frac{ip_3} {|y+p_2|^3}\right] dy \label{eq:parameters}\ee where
the independent parameters are
\begin{eqnarray} p_1 &=& 2\pi n \label{eq:p1}\\
p_2 &=& \frac{1}{2}\frac{w}{d} \label{eq:p2}\\
p_3 &=& \frac{C_3 \ell}{\hbar v d^3} .\label{eq:p3}
\end{eqnarray}
This exact choice of parameters is arbitrary, but convenient to
simplify equation \ref{eq:parameters}. The diffraction
efficiencies thus depend on the vdW coefficient, atom velocity,
grating period, grating thickness, and the grating open fraction.
This can be compared to equation \ref{eq:sinc} in which the
efficiencies depend only on the open fraction. If $C_3 = 0$, the
efficiencies depend only on $p_1$ and $p_2$, and equation
\ref{eq:parameters} reduces to equation \ref{eq:sinc}.

\begin{figure}[t]
\begin{center}
\includegraphics[width=8cm] {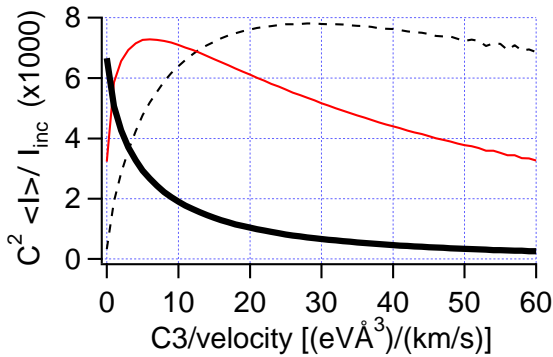}
\caption{The figure of merit $C^2\langle I \rangle / I_{inc}$ vs.
the vdW coefficient $C_3$ as calculated by equations \ref{eq:FOMeew}
and \ref{envdW}. Predictions are shown for gratings that have a
period of $d=100$ nm, and a thickness of $\ell = 150$ nm and
different sets of open fractions: (0.56, 0.50, 0.37) in thick solid
curve, (0.75, 0.67, 0.37) in thin solid curve and (0.93, 0.88, 0.37)
dashed curve. These open fractions were chosen because they maximize
$C^2 \langle I \rangle$ for various values of $C_3$ as shown in
Table \ref{tab:C3}.  The atom beam velocity is assumed to be
$v=1000$ m/s. \label{fig:FOM}}
\end{center}
\end{figure}

Now we have derived all the relationships (equations \ref{eq:FOMeew}
and \ref{eq:parameters}) needed to compute the figure of merit $C^2
\langle I \rangle$ as a function of the vdW coefficient $C_3$.
Figure \ref{fig:FOM} shows the quantity $C^2 \langle I \rangle/
I_{inc}$ as a function of $C_3$ for parameters ($v$=1000 m/s,
$d$=100 nm, \mbox{$\ell$ = 150 nm} and open fractions $(0.56, 0.50,
0.37)$) that are similar to those in experiments
\cite{ESC95,SCE95,RCK02,PEC05,HCL97,LHS97,TOE05} with supersonic
atom beams and state-of-the-art silicon nitride nanostructure
gratings. As shown in figure \ref{fig:FOM}, the figure of merit for
this set of parameters is a monotonically decreasing function of
$C_3$. In other words, the sensitivity to phase shifts becomes worse
as vdW interactions get stronger. For gratings with geometric
dimensions that are related by the ratios \mbox{($w:d:\ell =
1:2:3$)} similar to the 100-nm period gratings that are now
available, the quantity $C^2 \langle I \rangle / I_{inc}$ is reduced
by more than 1/2 (which means that a measurement of $\phi$ with
comparable uncertainty would require more than twice as much time)
when \be p_3 = \frac{C_3 \ell}{\hbar v d^3} > \frac{(5
\mbox{meVnm}^3) (150 \mbox{nm})}{\hbar (1000 \mbox{m/s}) (100
\mbox{nm})^3} = 1.1 \times 10^{-3} \label{eq:optimum parameter}.\ee
Since this condition is nearly satisfied in experiments
\cite{ESC95,SCE95,RCK02,PEC05,HCL97,LHS97,TOE05}, we are motivated
to ask the question, ``is it ever worth using sub-100-nm period
gratings for atom interferometry?" Or, to paraphrase Richard
Feynmann, ``is there no more room at the bottom?"  We will address
this question by exploring how the open fractions and all the
parameters in $p_3$ affect the figure of merit for minimum
$\sigma_{\phi}$;  then as a separate issue we will investigate how
the sensitivity to rotation and acceleration is affected by vdW
interactions.

The open fractions can be made larger or smaller by altering the
nanostructure fabrication procedure described in
\cite{SAS90,sss95,sava96}. So we used equations \ref{eq:FOMeew}
and \ref{eq:parameters} to specify what open fractions should be
chosen to optimize the interferometer if $C_3$ = 3 meV nm$^3$ and
the other parameters are kept \mbox{$v= 1000$ m/s}, $\ell= 150$
nm, and $d=100$ nm. The three terms that contribute to the result
for $C^2\langle I \rangle$ in equation \ref{eq:FOMeew} are each
plotted in figure \ref{fig:w123} as a function of open fraction
for the cases $C_3$ = 0,  3, and 30 meVnm$^3$.  The optimum open
fractions for each of these cases are summarized in table
\ref{tab:C3}, and were used to generate additional functions
plotted in figure \ref{fig:FOM} for $C^2 \langle I \rangle /
I_{inc}$ vs $C_3$.

The new values of $w_1$ and $w_2$ that maximize $C^2\langle I
\rangle$ when $C_3 \neq 0$ are significantly larger than the
optimum values found for the case of $C_3 = 0$.  This can be
understood qualitatively because larger open fractions are needed
to compensate for the effect of increased $C_3$;   increasing the
strength of the vdW interaction causes a change to $e_n$ that is
similar to (but not exactly the same as) the effect of decreasing
the open fraction of the grating.

If extremely large open fractions ($w \approx d$) are possible to
obtain for G1 and G2, then it is approximately correct to replace
the equation \ref{eq:optimum parameter}  by stating that $C^2
\langle I \rangle $ is reduced below 1/2 of its optimum value when
\be \frac{C_3 \ell}{\hbar v (w^3)}
> \frac{(5 \mbox{meVnm}^3) (150 \mbox{nm})}{\hbar (1000
\mbox{m/s}) (50 \mbox{nm})^3} = 9.1 \times 10^{-3} \label{eq:optimum
parameter_w}.\ee  This last condition is consistent with replacing
\mbox{$d\rightarrow 2w$} in equation \ref{eq:optimum parameter}, and
is verified by simulations. For grating-atom combinations that do
not satisfy the inequality \ref{eq:optimum parameter_w}, it is not
possible to restore $C^2 \langle I \rangle / I_{inc}$ to the optimum
value of 0.0070 regardless of the open fractions. This shows that
vdW interactions set a fundamental limit on the smallest
nanostructures that can be used for atom interferometry before the
performance as determined by the minimum $\sigma_{\phi}$ degrades.

Equation \ref{eq:optimum parameter_w} can be approximately derived
analytically by finding the condition for which the phase begins
to oscillate rapidly ($|\partial \phi(\xi) / \partial \xi|
> \pi / w$) when $\xi = d/4$. This analytic approach is justified
because the regions of rapidly oscillating phase do not contribute
significantly to the integral in equation \ref{envdW}, and we know
from our earlier study of optimization without vdW interactions
that the combination of replacing the limits by approximately
$-d/4$ and $d/4$ and ignoring the $\phi(\xi)$ term  in equation
\ref{envdW} yields the optimum $e_n$ for G1 and G2. The analytic
approach described here gives the condition \be \frac{C_3
\ell}{\hbar v} \left[\frac{w}{(d-2w)^4}\right]
> \left(\frac{\pi}{3}\frac{1}{4^4}\right) \approx 4 \times 10^{-3} \label{eq:optimum
parameter_w analytic} \ee which is consistent with equation
\ref{eq:optimum parameter_w} within a factor of $\pi$ for the case
$d \approx w$.

\begin{table}[t]
\caption{\label{tab:C3} Figures of merit tabulated for various open
fractions given different values of $C_3$ (in units of meVnm$^3$)
assuming the grating period $d$ = 100 nm grating thickness
$\ell$=150 nm and the atom velocity  $v=1000$ m/s.}
\begin{ruledtabular}
\begin{tabular}{llllccl}
$C_3$ &  $w_1/d$ &  $w_2/d$ &  $w_3/d$ & $C$ & $\langle I \rangle / I_{inc}$  & $C^2 \langle I \rangle /I_{inc}$\\
\hline
  0 &  0.56  & 0.50  & 0.37 &  0.67  &  0.015  &  0.0070  $\bigstar$\\
  3 &  0.56  & 0.50  & 0.37 &  0.76  &  0.007  &  0.0040    \\
 30 & 0.56  & 0.50  & 0.37 &   0.79  &  0.001  &  0.0008    \\  \hline
  3 &  0.75  & 0.67  & 0.37 &  0.66  &  0.016  &  0.0072  $\bigstar$\\
 30 &  0.93  & 0.88  & 0.37 &  0.68  &  0.016  &  0.0075  $\bigstar$\\
\end{tabular}
\end{ruledtabular}
\noindent $\bigstar$ optimized open fractions for the specific
$C_3$.
\end{table}

Identifying the independent parameter $p_3$ helps to clarify
several scaling laws.  For example, if every geometric dimension
of the gratings ($d$,$w$, and $\ell$) were multiplied by $1/a$,
then the efficiencies $e_n$ would change as if $C_3 \rightarrow
a^2 C_3$ because $\ell d^{-3}$ appears in $p_3$. For example, if
$\frac{1}{3}$-scale gratings could be obtained then the results in
figures \ref{fig:w123} and \ref{fig:FOM} for $C_3 = 27$ meVnm$^3$
would apply.

The vdW coefficient for He atoms is the smallest of any atom, so
the impact of vdW interactions should be minimal for He atoms. In
another example of a scaling law, the efficiencies $e_n$, and
figure of merit $C^2\langle I \rangle$ that we have discussed for
Na atoms (with $C_3$) will apply for the case of He atoms (with
$C_3^{\prime}$) if geometrically similar gratings with new period
of $d^{\prime} = (C^{\prime}_3/C_3)^{1/2}d \approx 0.18 d$ are
used. Thus, for a He atom beam with $v=1000$ m/s, nanostructure
gratings with a minimum dimension of $d \approx 2w = 16$ nm could
be used before $\sigma^2_{\phi}$ is doubled from vdW interactions.

Next we address  the question, ``What period grating would
optimize an atom beam gyrometer or accelerometer?" Because the
interference fringe phase shift, $\phi$, due to either rotation or
acceleration depends on $d$ and $v$ there is a different function
to optimize for optimum sensitivity to inertial displacements.

The Sagnac phase shift for a matter wave interferometer rotating
at the rate $\Omega$ is
\begin{eqnarray} \phi &=& 4 \pi (\vec{\Omega} \cdot \vec{A}) \frac
{m}{h}\\ &=& \frac{4 \pi \Omega L^2}{d v} \end{eqnarray} where $A$
is the area enclosed by the interferometer paths, and $L$ is the
distance between gratings G1 and G2 (or equivalently G2 and G3)
\cite{LHS97}. In the last equation it was assumed that the vector
orientations of $\Omega$ and $A$ are parallel. The statistical
variance in measured rotation rate will then be given by
\begin{eqnarray} (\sigma_{\Omega})^2 &=& (\frac{\partial
\Omega}{\partial \phi} \sigma_{\phi})^2 \\ &=& (\frac{dv}{4 \pi L^2}
\frac{1}{C \sqrt{N}})^2.
\end{eqnarray} Hence, maximizing the quantity \be F = \frac{(4\pi)^2L^4}{d^2v^2}\frac{C^2
\langle I \rangle }{I_{inc}} \label{eq:F}\ee will minimize the
variance in measured rotation rate, and will therefore minimize the
angle random walk obtained when using a gyrometer for inertial
navigation.  For measurements of acceleration with minimum variance,
the quantity $F(2v)^{-2}$ should be maximized.

\begin{figure}[t]
\begin{center}
\includegraphics[width=8cm] {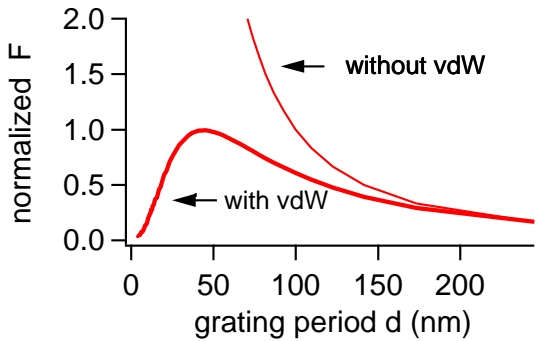}
\caption{The figure of merit for inertial sensors, $F$, given by
equation \ref{eq:F} is plotted vs grating period given the
restriction $w:d:\ell= 1:2:3$ for all three gratings, and the
parameters $C_3 = 3$ meVnm$^3$, and $v=1000$ m/s (thick line). For
comparison if $C_3 = 0$ (thin line) then $F$ depends on
$d^{-2}$.\label{fig:inertialFOM}}
\end{center}
\end{figure}

The tradeoff is as follows.  For smaller $d$ an atom
interferometer gyroscope is limited by vdW interactions, and for
larger $d$ the response factor ($d\phi/d\Omega$) is smaller. The
quantity $F$ is plotted in figure \ref{fig:inertialFOM} as a
function of $d$ assuming $v=1000$ m/s, $C_3 = 3$ meVnm$^3$ and
\mbox{($w:d:\ell =1:2:3$)} for each grating.  An optimum period of
$d=$  44 nm is found.

The main result is that nanostructure gratings with a period
smaller than $d = 44$ nm will not improve atom beam interferometer
gyroscopes or accelerometers unless the vdW limitation is
overcome. This limitation can be circumvented somewhat if large
open fractions are used, or by choosing atom-surface combinations
that have a small vdW coefficient $C_3$, or by reducing the
grating thickness $\ell$ independently of $d$ as shown in equation
\ref{eq:optimum parameter} or by adjusting atom velocity to
maximize the figure of merit shown in equation \ref{eq:F}. Some of
these advantages could be realized if gratings were fabricated
from an array of carbon nanotubes.

In conclusion, we have shown how atom interferometers are affected
by vdW interactions between atoms and nanostructure gratings. We
have shown how to calculate the contrast and intensity of the
interference pattern given the parameters: grating thickness
($\ell$), grating period ($d$), grating open fractions ($w_i/d)$,
atom velocity ($v$), and vdW coefficient ($C_3$).   We described
how to select open fractions that will optimize an interferometer
for maximum sensitivity to interference fringe phase shifts.  For
experiments with the currently available parameters ($\ell$ = 150
nm, $d$= 100 nm, $v$ = 1000 m/s, $C_3$ = 3 meVnm$^3$), we report
the open fractions that minimize the uncertainty in phase
($\sigma_{\phi}$) are given by
$(w_1/d,w_2/d,w_3/d)=(0.75,0.67,0.37)$. If the gratings are made
with a period smaller than $d=50$ nm, then regardless of the open
fractions, the statistical sensitivity to phase shifts (given by
the uncertainty $\sigma_{\phi}$) will grow because of the way vdW
interactions adversely affect how the gratings operate as beam
splitters. For maximum sensitivity to rotation or acceleration,
there is a minimum grating period in the range of 40 nm.  Thus van
der Waals interactions place a limitation on the smallest
nanostructure gratings that can be used for atom interferometry.

This work was supported by an award from Research Corporation and
the National Science Foundation Grant No PHY-0354947.

%\bibliography{opt_references}

\begin{thebibliography}{22}
\expandafter\ifx\csname
natexlab\endcsname\relax\def\natexlab#1{#1}\fi
\expandafter\ifx\csname bibnamefont\endcsname\relax
  \def\bibnamefont#1{#1}\fi
\expandafter\ifx\csname bibfnamefont\endcsname\relax
  \def\bibfnamefont#1{#1}\fi
\expandafter\ifx\csname citenamefont\endcsname\relax
  \def\citenamefont#1{#1}\fi
\expandafter\ifx\csname url\endcsname\relax
  \def\url#1{\texttt{#1}}\fi
\expandafter\ifx\csname
urlprefix\endcsname\relax\def\urlprefix{URL }\fi
\providecommand{\bibinfo}[2]{#2}
\providecommand{\eprint}[2][]{\url{#2}}

\bibitem[{\citenamefont{Ekstrom et~al.}(1995)\citenamefont{Ekstrom,
  Schmiedmayer, Chapman, Hammond, and Pritchard}}]{ESC95}
\bibinfo{author}{\bibfnamefont{C.}~\bibnamefont{Ekstrom}},
  \bibinfo{author}{\bibfnamefont{J.}~\bibnamefont{Schmiedmayer}},
  \bibinfo{author}{\bibfnamefont{M.}~\bibnamefont{Chapman}},
  \bibinfo{author}{\bibfnamefont{T.}~\bibnamefont{Hammond}}, \bibnamefont{and}
  \bibinfo{author}{\bibfnamefont{D.~E.} \bibnamefont{Pritchard}},
  \bibinfo{journal}{Phys. Rev. A} \textbf{\bibinfo{volume}{51}},
  \bibinfo{pages}{3883} (\bibinfo{year}{1995}).

\bibitem[{\citenamefont{Schmiedmayer et~al.}(1995)\citenamefont{Schmiedmayer,
  Chapman, Ekstrom, Hammond, Wehinger, and Pritchard}}]{SCE95}
\bibinfo{author}{\bibfnamefont{J.}~\bibnamefont{Schmiedmayer}},
  \bibinfo{author}{\bibfnamefont{M.}~\bibnamefont{Chapman}},
  \bibinfo{author}{\bibfnamefont{C.}~\bibnamefont{Ekstrom}},
  \bibinfo{author}{\bibfnamefont{T.}~\bibnamefont{Hammond}},
  \bibinfo{author}{\bibfnamefont{S.}~\bibnamefont{Wehinger}}, \bibnamefont{and}
  \bibinfo{author}{\bibfnamefont{D.}~\bibnamefont{Pritchard}},
  \bibinfo{journal}{Phys. Rev. Lett.} \textbf{\bibinfo{volume}{74}},
  \bibinfo{pages}{1043} (\bibinfo{year}{1995}).

\bibitem[{\citenamefont{Roberts et~al.}(2002)\citenamefont{Roberts, Cronin,
  Kokorowski, and Pritchard}}]{RCK02}
\bibinfo{author}{\bibfnamefont{T.~D.} \bibnamefont{Roberts}},
  \bibinfo{author}{\bibfnamefont{A.~D.} \bibnamefont{Cronin}},
  \bibinfo{author}{\bibfnamefont{D.~A.} \bibnamefont{Kokorowski}},
  \bibnamefont{and} \bibinfo{author}{\bibfnamefont{D.~E.}
  \bibnamefont{Pritchard}}, \bibinfo{journal}{Physical Review Letters}
  \textbf{\bibinfo{volume}{89}} (\bibinfo{year}{2002}).

\bibitem[{\citenamefont{Perreault and Cronin}(2005)}]{PEC05}
\bibinfo{author}{\bibfnamefont{J.~D.} \bibnamefont{Perreault}}
  \bibnamefont{and} \bibinfo{author}{\bibfnamefont{A.~D.}
  \bibnamefont{Cronin}}, \bibinfo{journal}{arXiv:physics/0505160}
  (\bibinfo{year}{2005}).

\bibitem[{\citenamefont{Hammond et~al.}(1997)\citenamefont{Hammond, Chapman,
  Lenef, Schmiedmayer, Smith, Rubenstein, Kokorowski, and Pritchard}}]{HCL97}
\bibinfo{author}{\bibfnamefont{T.}~\bibnamefont{Hammond}},
  \bibinfo{author}{\bibfnamefont{M.}~\bibnamefont{Chapman}},
  \bibinfo{author}{\bibfnamefont{A.}~\bibnamefont{Lenef}},
  \bibinfo{author}{\bibfnamefont{J.}~\bibnamefont{Schmiedmayer}},
  \bibinfo{author}{\bibfnamefont{E.}~\bibnamefont{Smith}},
  \bibinfo{author}{\bibfnamefont{R.}~\bibnamefont{Rubenstein}},
  \bibinfo{author}{\bibfnamefont{D.}~\bibnamefont{Kokorowski}},
  \bibnamefont{and}
  \bibinfo{author}{\bibfnamefont{D.}~\bibnamefont{Pritchard}},
  \bibinfo{journal}{Braz. J. Phys.} \textbf{\bibinfo{volume}{27}},
  \bibinfo{pages}{193} (\bibinfo{year}{1997}).

\bibitem[{\citenamefont{Lenef et~al.}(1997)\citenamefont{Lenef, Hammond, Smith,
  Chapman, Rubenstein, and Pritchard}}]{LHS97}
\bibinfo{author}{\bibfnamefont{A.}~\bibnamefont{Lenef}},
  \bibinfo{author}{\bibfnamefont{T.}~\bibnamefont{Hammond}},
  \bibinfo{author}{\bibfnamefont{E.}~\bibnamefont{Smith}},
  \bibinfo{author}{\bibfnamefont{M.}~\bibnamefont{Chapman}},
  \bibinfo{author}{\bibfnamefont{R.}~\bibnamefont{Rubenstein}},
  \bibnamefont{and}
  \bibinfo{author}{\bibfnamefont{D.}~\bibnamefont{Pritchard}},
  \bibinfo{journal}{Phys. Rev Lett.} \textbf{\bibinfo{volume}{78}}
  (\bibinfo{year}{1997}).

\bibitem[{\citenamefont{Scully and Dowling}(1993)}]{SCD93}
\bibinfo{author}{\bibfnamefont{M.}~\bibnamefont{Scully}} \bibnamefont{and}
  \bibinfo{author}{\bibfnamefont{J.}~\bibnamefont{Dowling}},
  \bibinfo{journal}{Phys. Rev. A} \textbf{\bibinfo{volume}{48}},
  \bibinfo{pages}{3186} (\bibinfo{year}{1993}).

\bibitem[{\citenamefont{Berman}(1997)}]{BER97}
\bibinfo{editor}{\bibfnamefont{P.~R.} \bibnamefont{Berman}}, ed.,
  \emph{\bibinfo{title}{Atom Interferometry}} (\bibinfo{publisher}{Academic
  Press}, \bibinfo{year}{1997}).

\bibitem[{\citenamefont{Champenois et~al.}(1999)\citenamefont{Champenois,
  Buchner, and Vigue}}]{CBV99}
\bibinfo{author}{\bibfnamefont{C.}~\bibnamefont{Champenois}},
  \bibinfo{author}{\bibfnamefont{M.}~\bibnamefont{Buchner}}, \bibnamefont{and}
  \bibinfo{author}{\bibfnamefont{J.}~\bibnamefont{Vigue}},
  \bibinfo{journal}{European Physical Journal D} \textbf{\bibinfo{volume}{5}},
  \bibinfo{pages}{363} (\bibinfo{year}{1999}).

\bibitem[{\citenamefont{Delhuille et~al.}(2002)\citenamefont{Delhuille, Miffre,
  de~Lesegno, Buchner, Rizzo, Trenec, and Vigue}}]{DML02}
\bibinfo{author}{\bibfnamefont{R.}~\bibnamefont{Delhuille}},
  \bibinfo{author}{\bibfnamefont{A.}~\bibnamefont{Miffre}},
  \bibinfo{author}{\bibfnamefont{B.~V.} \bibnamefont{de~Lesegno}},
  \bibinfo{author}{\bibfnamefont{M.}~\bibnamefont{Buchner}},
  \bibinfo{author}{\bibfnamefont{C.}~\bibnamefont{Rizzo}},
  \bibinfo{author}{\bibfnamefont{G.}~\bibnamefont{Trenec}}, \bibnamefont{and}
  \bibinfo{author}{\bibfnamefont{J.}~\bibnamefont{Vigue}},
  \bibinfo{journal}{Acta Physica Polonica B} \textbf{\bibinfo{volume}{33}},
  \bibinfo{pages}{2157} (\bibinfo{year}{2002}).

\bibitem[{\citenamefont{Grisenti et~al.}(1999)\citenamefont{Grisenti,
  Schollkopf, Toennies, Hegerfeldt, and Kohler}}]{gris99}
\bibinfo{author}{\bibfnamefont{R.~E.} \bibnamefont{Grisenti}},
  \bibinfo{author}{\bibfnamefont{W.}~\bibnamefont{Schollkopf}},
  \bibinfo{author}{\bibfnamefont{J.~P.} \bibnamefont{Toennies}},
  \bibinfo{author}{\bibfnamefont{G.~C.} \bibnamefont{Hegerfeldt}},
  \bibnamefont{and} \bibinfo{author}{\bibfnamefont{T.}~\bibnamefont{Kohler}},
  \bibinfo{journal}{Phys. Rev. Lett.} \textbf{\bibinfo{volume}{83}},
  \bibinfo{pages}{1755} (\bibinfo{year}{1999}).

\bibitem[{\citenamefont{Bruhl et~al.}(2002)\citenamefont{Bruhl, Fouquet,
  Grisenti, Toennies, Hegerfeldt, Kohler, Stoll, and Walter}}]{bruh02}
\bibinfo{author}{\bibfnamefont{R.}~\bibnamefont{Bruhl}},
  \bibinfo{author}{\bibfnamefont{P.}~\bibnamefont{Fouquet}},
  \bibinfo{author}{\bibfnamefont{R.~E.} \bibnamefont{Grisenti}},
  \bibinfo{author}{\bibfnamefont{J.~P.} \bibnamefont{Toennies}},
  \bibinfo{author}{\bibfnamefont{G.~C.} \bibnamefont{Hegerfeldt}},
  \bibinfo{author}{\bibfnamefont{T.}~\bibnamefont{Kohler}},
  \bibinfo{author}{\bibfnamefont{M.}~\bibnamefont{Stoll}}, \bibnamefont{and}
  \bibinfo{author}{\bibfnamefont{D.}~\bibnamefont{Walter}},
  \bibinfo{journal}{Europhys. Lett.} \textbf{\bibinfo{volume}{59}},
  \bibinfo{pages}{357} (\bibinfo{year}{2002}).

\bibitem[{\citenamefont{Grisenti et~al.}(2000)\citenamefont{Grisenti,
  Schollkopf, Toennies, Manson, Savas, and Smith}}]{GST00}
\bibinfo{author}{\bibfnamefont{R.~E.} \bibnamefont{Grisenti}},
  \bibinfo{author}{\bibfnamefont{W.}~\bibnamefont{Schollkopf}},
  \bibinfo{author}{\bibfnamefont{J.~P.} \bibnamefont{Toennies}},
  \bibinfo{author}{\bibfnamefont{J.~R.} \bibnamefont{Manson}},
  \bibinfo{author}{\bibfnamefont{T.~A.} \bibnamefont{Savas}}, \bibnamefont{and}
  \bibinfo{author}{\bibfnamefont{H.~I.} \bibnamefont{Smith}},
  \bibinfo{journal}{Phys. Rev. A} \textbf{\bibinfo{volume}{61}},
  \bibinfo{pages}{033608} (\bibinfo{year}{2000}).

\bibitem[{\citenamefont{Perreault et~al.}(2005)\citenamefont{Perreault, Cronin,
  and Savas}}]{PCS05}
\bibinfo{author}{\bibfnamefont{J.~D.} \bibnamefont{Perreault}},
  \bibinfo{author}{\bibfnamefont{A.~D.} \bibnamefont{Cronin}},
  \bibnamefont{and} \bibinfo{author}{\bibfnamefont{T.~A.} \bibnamefont{Savas}},
  \bibinfo{journal}{Phys. Rev. A} \textbf{\bibinfo{volume}{71}},
  \bibinfo{pages}{053612} (\bibinfo{year}{2005}).

\bibitem[{\citenamefont{Cronin and Perreault}(2004)}]{CRP04}
\bibinfo{author}{\bibfnamefont{A.~D.} \bibnamefont{Cronin}} \bibnamefont{and}
  \bibinfo{author}{\bibfnamefont{J.~D.} \bibnamefont{Perreault}},
  \bibinfo{journal}{Phys. Rev. A} \textbf{\bibinfo{volume}{70}},
  \bibinfo{pages}{043607} (\bibinfo{year}{2004}).

\bibitem[{\citenamefont{Keith et~al.}(1991)\citenamefont{Keith, Ekstrom,
  Turchette, and Pritchard}}]{KET91}
\bibinfo{author}{\bibfnamefont{D.~W.} \bibnamefont{Keith}},
  \bibinfo{author}{\bibfnamefont{C.~R.} \bibnamefont{Ekstrom}},
  \bibinfo{author}{\bibfnamefont{Q.~A.} \bibnamefont{Turchette}},
  \bibnamefont{and}
  \bibinfo{author}{\bibfnamefont{D.}~\bibnamefont{Pritchard}},
  \bibinfo{journal}{Phys. Rev. Lett.} \textbf{\bibinfo{volume}{66}},
  \bibinfo{pages}{2693} (\bibinfo{year}{1991}).

\bibitem[{\citenamefont{Chapman and \textit{et~al.}}(1995)}]{CEH95}
\bibinfo{author}{\bibfnamefont{M.~S.} \bibnamefont{Chapman}} \bibnamefont{and}
  \bibinfo{author}{\bibnamefont{\textit{et~al.}}}, \bibinfo{journal}{Phys. Rev.
  Lett.} \textbf{\bibinfo{volume}{74}}, \bibinfo{pages}{4783}
  (\bibinfo{year}{1995}).

\bibitem[{\citenamefont{Toennies}(2001)}]{TOE01}
\bibinfo{author}{\bibfnamefont{J.~P.} \bibnamefont{Toennies}},
  \bibinfo{journal}{Hinshelwood Lecutures, Oxford}  (\bibinfo{year}{2001}).

\bibitem[{\citenamefont{Toennies}(2005)}]{TOE05}
\bibinfo{author}{\bibfnamefont{J.~P.} \bibnamefont{Toennies}},
  \bibinfo{journal}{personal commuinication}  (\bibinfo{year}{2005}).

\bibitem[{\citenamefont{Schattenburg et~al.}(1990)\citenamefont{Schattenburg,
  Anderson, and Smith}}]{SAS90}
\bibinfo{author}{\bibfnamefont{M.~l.} \bibnamefont{Schattenburg}},
  \bibinfo{author}{\bibfnamefont{E.~H.} \bibnamefont{Anderson}},
  \bibnamefont{and} \bibinfo{author}{\bibfnamefont{H.~I.} \bibnamefont{Smith}},
  \bibinfo{journal}{Phys. Scripta} \textbf{\bibinfo{volume}{41}},
  \bibinfo{pages}{13} (\bibinfo{year}{1990}).

\bibitem[{\citenamefont{Savas et~al.}(1995)\citenamefont{Savas, Shah,
  Schattenburg, Carter, and Smith}}]{sss95}
\bibinfo{author}{\bibfnamefont{T.~A.} \bibnamefont{Savas}},
  \bibinfo{author}{\bibfnamefont{S.~N.} \bibnamefont{Shah}},
  \bibinfo{author}{\bibfnamefont{M.~L.} \bibnamefont{Schattenburg}},
  \bibinfo{author}{\bibfnamefont{J.~M.} \bibnamefont{Carter}},
  \bibnamefont{and} \bibinfo{author}{\bibfnamefont{H.~I.} \bibnamefont{Smith}},
  \bibinfo{journal}{Journal of Vacuum Science and Technology B}
  \textbf{\bibinfo{volume}{13}}, \bibinfo{pages}{2732} (\bibinfo{year}{1995}).

\bibitem[{\citenamefont{Savas et~al.}(1996)\citenamefont{Savas, Schattenburg,
  Carter, and Smith}}]{sava96}
\bibinfo{author}{\bibfnamefont{T.~A.} \bibnamefont{Savas}},
  \bibinfo{author}{\bibfnamefont{M.~L.} \bibnamefont{Schattenburg}},
  \bibinfo{author}{\bibfnamefont{J.~M.} \bibnamefont{Carter}},
  \bibnamefont{and} \bibinfo{author}{\bibfnamefont{H.~I.} \bibnamefont{Smith}},
  \bibinfo{journal}{J. Vac. Sci. Tech. B} \textbf{\bibinfo{volume}{14}},
  \bibinfo{pages}{4167} (\bibinfo{year}{1996}).

\end{thebibliography}

\end{document}